# Neural Network-based Equalizer by Utilizing Coding Gain in Advance

Chieh-Fang Teng[1*], Han-Mo Ou[2*], An-Yeu (Andy) Wu[1], *Fellow*, *IEEE*
[1]Graduate Institute of Electrical Engineering, National Taiwan University, Taipei, Taiwan
[2]Department of Electrical Engineering, National Taiwan University, Taipei, Taiwan
jeff@access.ee.ntu.edu.tw, b05901092@ntu.edu.tw, andywu@ntu.edu.tw

*Abstract*—Recently, deep learning has been exploited in many fields with revolutionary breakthroughs. In the light of this, deep learning-assisted communication systems have also attracted much attention in recent years and have potential to break down the conventional design rule for communication systems. In this work, we propose two kinds of neural network-based equalizers to exploit different characteristics between convolutional neural networks and recurrent neural networks. The equalizer in conventional block-based design may destroy the code structure and degrade the capacity of coding gain for decoder. On the contrary, our proposed approach not only eliminates channel fading, but also exploits the code structure with utilization of coding gain in advance, which can effectively increase the overall utilization of coding gain with more than 1.5 dB gain.

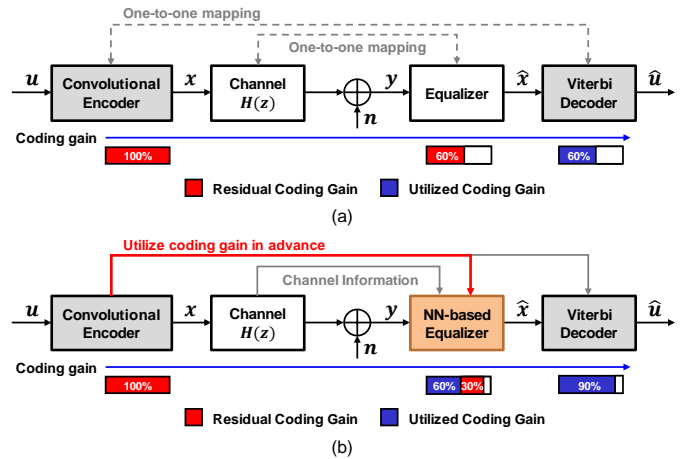

**Fig. 1.** Overview of (a) conventional communication system with functional blocks designed in one-to-one mapping, and (b) proposed neural network-based equalizer with utilization of coding gain in advance.

## I. INTRODUCTION

With more and more revolutionary breakthroughs in the fields of computer vision and natural language processing, machine learning-assisted communication systems have gradually attracted attention in recent years [1]-[2]. For example, a powerful modulation classifier is proposed in [3]-[5] by taking advantages of convolutional neural networks. For channel decoding, the authors in [6]-[8] propose neural network-based decoders for BCH code and polar code, which have better performance and faster convergence. Furthermore, under complicated channel interference, signal detection in orthogonal frequency-division multiplexing (OFDM) and multiple-input-multiple-output (MIMO) systems are proposed to achieve state-of-the-art accuracies with significantly lower complexity while providing robustness [9]-[10]. Based on above fruitful results, deep learning (DL) has great potential and is promising for overcoming the more complicated scenario and meeting the rigorous requirements in future communications.

In this work, we mainly expect to convey a brand-new concept for communication systems by taking advantages of DL. The overview of a conventional communication system is shown in Fig. 1(a) with functional blocks designed in one-to-one mapping. Namely, the decoder is designed based on the encoder to exploit coding gain and the equalizer is planned to eliminate channel fading. Here, coding gain represents the capacity of error correction benefiting from the code. This approach has been adopted and led to the efficient and controllable systems we have today. However, it also constrains the development and causes long-standing defects in our communication systems. Because coding gain can only be exploited by the decoder in the last stage, functional blocks between the encoder and decoder, such as the equalizer, destroy the code structure and degrade the capacity of coding gain as shown in Fig. 1(a).

To address this issue, neural network-based equalizers are proposed to simultaneously eliminate channel fading and exploit code structure from received signal, which can effectively increase the overall utilization of coding gain with better performance. Fig. 1(b) shows an overview of the proposed system, and our main contributions are summarized as follows:

1) By exploiting different model architectures, convolutional neural network-based and recurrent neural network-based equalizers are proposed with detailed evaluation. The proposed system can outperform conventional approaches by 0.6 dB and 1.5 dB by utilizing coding gain in advance.

2) A brand-new concept is proposed to break down the conventional design rule for communication systems, which provides highly promising performances and has the potential to trigger more interesting research in this direction.

The rest of this paper is organized as follows. Section II briefly reviews the system model and conventional equalizers. Section III illustrates the architecture of proposed neural network-based equalizers. The numerical experiments and analyses are shown in Section IV. Finally, Section V concludes this paper.

[*]These two authors contributed equally.

This research work is financially supported by the MediaTek Inc., Hsinchu, Taiwan, under Grants MTKC-2019-0070. The first author is also sponsored by MediaTek Ph.D. Fellowship program.

## II. BACKGROUND

### A. System Model

The system model is depicted in Fig. 1. At the transmitter side, the information messages $\boldsymbol{u}$ are first encoded as $\boldsymbol{x}$ by the convolutional encoder and then sent to the channel. Throughout this paper, channel fading is considered. Therefore, inter-symbol interference (ISI) and additive white Gaussian noise (AWGN) jointly contribute to channel distortion and the received signal can be expressed as:

$$y[i] = \sum_{j=0}^{L-1} x[i-j] \times h[j] + n[i], \quad (1)$$

where $L$ is the length of the response, $\boldsymbol{h}$ and $\boldsymbol{n}$ are the channel impulse response and Gaussian noise, respectively.

At the receiver side, the equalizer is firstly applied to eliminate the ISI effects, which will be reviewed in the Section II.B. Then, based on the reconstructed signal $\hat{\boldsymbol{x}}$ from the equalizer, a Viterbi decoder is used to estimate $\hat{\boldsymbol{u}}$ for the original information messages $\boldsymbol{u}$. Thus, the characteristic of one-to-one mapping in communication systems is obvious with the equalizer and Viterbi decoder corresponding to channel fading and the convolutional encoder, respectively. For this reason, the equalizer is designed without considering the code structure and results in degradation of coding gain.

### B. Prior Works of Channel Equalization

Inter-symbol interference occurs when the transmitted signal has multiple paths to reach the receiver, which results in combination of symbols over time with severe interference. The process of removing the ISI is called equalization and least mean square (LMS) algorithm is one of the widely adopted adaptive filters for channel equalization. The main idea of LMS is to adapt a filter $\hat{\boldsymbol{h}}$ such that the convolution with received signal $\boldsymbol{y}$ is close to the training sequence $\boldsymbol{d}$ and can be expressed as:

$$e[i] = d[i] - \sum_{j=0}^{L-1} y[i-j] \times \hat{h}[j], \quad (2)$$

where $\boldsymbol{e}$ is the minimized target by gradient descent algorithm to obtain optimum filter weights.

Compared to the linear equalizer of LMS, neural network architectures are also exploited to deal with nonlinear distortion, which benefits from its nonlinear transformation function in deep architecture [11]-[13]. In [11], a multi-layer perceptron (MLP) structure is proposed, which has promising performance for nonlinear channels and colored noise. Furthermore, in [12]-[13], the authors propose a neural network for joint channel equalization and decoding, which represents that the network has the ability to simultaneously address various channel effects and learn the complicated decoder function. Besides, the end-to-end optimization manner also leads to a better performance.

However, in [11], the authors are only dedicated to reconstruct the signal without considering the code structure. On the other hand, the nonlinear channels in [12]-[13] are fixed, which means that neural networks can cram for the channel and outperform conventional approaches easily. To address above concerns, we only consider the linear ISI effect in this work. By decreasing the variables, the ability of proposed neural network-based equalizers can be carefully analyzed. For future extension of more complicated channels, we believe that the improvement can be even bigger compared to the simple channel used in this work.

## III. PROPOSED NEURAL NETWORK-BASED EQUALIZER

To address the aforementioned issues, inspired by LMS algorithm and the characteristics of received signals, we exploit different model architectures and propose convolutional neural network-based and recurrent neural network-based equalizers.

### A. Convolutional Neural Network-Based Equalizer

Convolutional neural networks (CNN) are commonly used in computer vision for feature extraction and image recognition. A convolutional layer applies a sliding window to extract subtle patterns on the input data. As the window slides, the resulting weighted sum is computed as the output, hence the name "convolution". The main concept in using deep CNNs is to regard large patterns as the combination of smaller patterns. A typical convolutional layer consists of a number of kernels, which extract different features from the inputs. Convolutional neural networks have lower complexity compared to fully connected networks, since the only trainable parameters are the kernel weights regardless of input size.

Considering the structure of convolutional neural networks, and taking into account that the convolutional sum is intrinsically installed in traditional linear filters, we propose that a CNN model be trained to perform as an equalizer. The proposed architecture of equalizer is illustrated in Fig. 2(b). The architecture of the CNN model is constructed by two one-dimensional convolutional layers and followed by a dense layer. The values below the convolutional layer represent the number and size of filters, and the values below the dense layer represent the number of nodes. The nonlinear function used by the two convolutional layers, Rectified Linear Units (ReLUs), is defined as:

$$f_{\text{ReLU}}(x) = \max\{0, x\} \quad (3)$$

The activation functions can result in a nonlinear system and augment the learning capability of the neural network, allowing it to outperform conventional linear filters. For the proposed CNN-based equalizer, six symbols are equalized at a time, while three additional symbols, so called "taps", are included at each end as shown in Fig. 2 (a), which can avoid the performance loss on the boundary. Therefore, the input data for the proposed equalizer contains twelve symbols with two channels representing the in-phase and quadrature components. The loss function for training is mean-square error, which measures the average squared distance between predicted positions and actual positions on the constellation map. Therefore, the predicted symbols would be trained to be close to the exact symbols on the constellation map.

### B. Recurrent Neural Network-Based Equalizer

Recurrent neural networks (RNN) are commonly used to analyze and predict time sequences, such as speech and video. With their internal memories, the output of RNNs would depend on not only the current input, but previous inputs as well. As ISI and convolutional code both impact the signal along the time axis, strengthening the relation between adjacent received symbols, we propose a recurrent neural network-based equalizer to exploit the relation between adjacent received symbols as shown in Fig. 2(c).

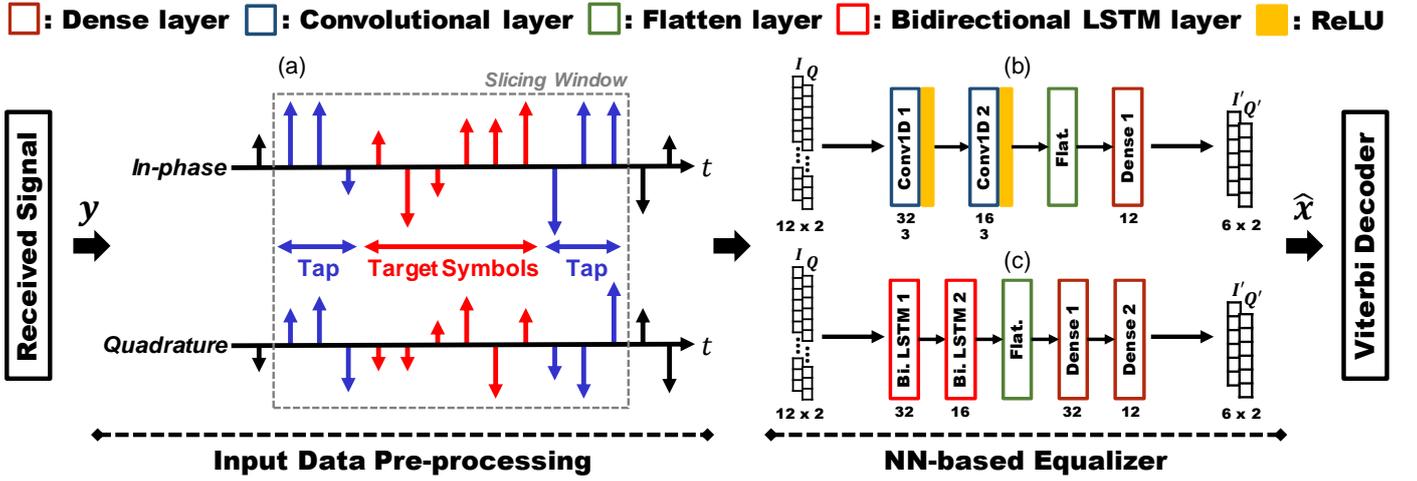

**Fig. 2.** The detailed overview of the receiver: (a) illustration of the process of input data pre-processing; (b) proposed convolutional neural network-based equalizer; (c) proposed recurrent neural network-based equalizer.

In the proposed architecture, the utilized long short-term memory (LSTM) layer is a class of RNN with each unit composed of a cell, an input, an output gate, and a forget gate [14]. The computation inside an LSTM unit at time $t$ can be expressed as:

$$i_t = \sigma(U_i x_t + V_i h_{t-1} + b_i) \quad (4)$$

$$o_t = \sigma(U_o x_t + V_o h_{t-1} + b_o) \quad (5)$$

$$f_t = \sigma(U_f x_t + V_f h_{t-1} + b_f) \quad (6)$$

$$c_t = i_t \circ \tanh(U_c x_t + V_c h_{t-1} + b_c) + f_t \circ c_{t-1} \quad (7)$$

$$h_t = o_t \circ \tanh(c_t) \quad (8)$$

where $i_t, o_t$ and $f_t$ represent input gate, output gate and forget gate activation vectors, respectively. $U, V$ and $b$ are trainable weights, $\sigma$ is logistic sigmoid function, and $\circ$ denotes point-wise multiplication. From (7) and (8), we can observe that the next state memories $c_t$ and output $h_t$ are determined by both the input $x_t$ and the previous hidden memories. Besides, the three gates are utilized to control the flow of information into and out of the cell and the duration of memory. Therefore, LSTM permits the analysis of sequential data, especially for detection along time series.

The proposed architecture features two bidirectional LSTM layers and two dense layers. The values below the bidirectional LSTM layer represent the number of cells. Bidirectional LSTM utilizes an additional LSTM layer in parallel to process the time series in reverse. We opt for such model structure to consider both forwards and backwards information when equalizing the received symbols. The input and output size are identical to the CNN model as shown in Fig. 2(a).

**TABLE I.** Simulation parameters.

| Modulation category | QPSK |
|---|---|
| Training data | 480000 bits |
| Testing data | 1920000 bits |
| Fading channel | $h(z) = 0.84 + 0.47z^{-1} + 0.28z^{-2}$ |
| Convolutional Code | (2, 1, 2) |
| Generator Polynomial | [3,7] |
| Signal to Noise Ratio (SNR) | 0, 2, 4, 6, 8 (dB) |
| Optimizer | SGD with Adam |
| Training and testing environment | Deep learning library of Keras running on top of TensorFlow with NVIDIA GTX 980 Ti GPU |

For the training process, the model input of our equalizer is the received symbols, while the target output is the transmitted symbols. Furthermore, 20% of the training data are used as the validation set with the mechanism of early stopping. Therefore, after each training epoch, the validation error is calculated and used to determine whether to terminate the training process.

IV. EXPERIMENTAL RESULTS AND ANALYSIS

As mentioned before, the channel considered in this paper is a fading channel with AWGN. Although we believe that other complex and nonlinear channel effects can be dealt with by NNs with superior performance against conventional approaches, we simplify the utilized channel to evaluate the ability of the proposed NN-based equalizers. The neural networks only take place of the equalizer while the other blocks remain unchanged throughout the experiment as shown in Fig. 1(b). The simulation setup is summarized in Table I.

*A. Simulation with Convolutional Code*

Convolutional codes are a class of error correction codes that transform the data sequence with a set of Boolean polynomials. A $(n, k, m)$ convolutional code has $k$ inputs, $n$ outputs, and $m$ memory units. Therefore, the encoder can be implemented in a shift register structure with a total of $m$ memory registers. In this paper, we adopt a convolutional code with code rate $1/2$, which indicates that the output sequence is twice as long as the input. Throughout our experiment, the generator polynomials remain unchanged as shown in Table I.

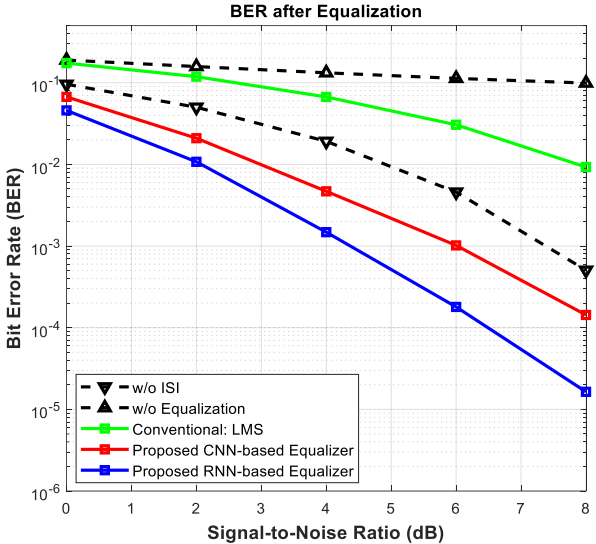
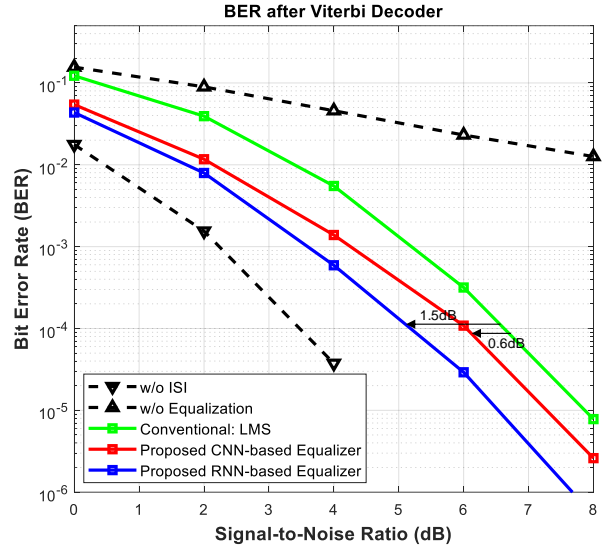

**Fig. 3.** Comparison of BER after equalization under different signal-to-noise ratio.

**Fig. 4.** Comparison of BER after Viterbi decoder under different signal-to-noise ratio.

In this experiment, the data bits are first passed to the convolutional encoder before modulation. Then, we compare the bit error rate (BER) before passing the Viterbi decoder, namely to evaluate the ability of signal reconstruction of equalizer under channel coding. In Fig. 3, we compare the BER of the proposed CNN-based and RNN-based equalizers with the conventional block-based LMS filter design. Besides, the BER of the worst case (w/o equalization) and the ideal case (w/o ISI) are also evaluated for exhaustive comparison.

Fig. 3 shows that proposed NN-based equalizers achieve significant accuracy improvements compared to conventional LMS. Furthermore, it can even outperform the ideal case, which is only distorted by AWGN. This informs us that the errors caused by AWGN, which is usually unfixable, have been removed by the NN-based equalizers. In other words, the proposed equalizer not only functions as an inverse to the channel response, but also utilizes the coding gain in advance, providing error correction simultaneously. Moreover, RNN is more suitable to exploit the relation between adjacent symbols with better BER than CNN.

*B. Simulation with Convolutional Code and Viterbi Decoder*

Viterbi algorithm is a dynamic programming algorithm used to find the path of state transition with the maximum likelihood. In the case of convolutional code decoding, it is the most commonly used method to predict the original bit sequence. Soft Viterbi algorithm is used in our experiments, which requires the soft input of each symbol without digitalization. Therefore, we treat the output of the equalizers as the coordination on the map for further processing and decoding.

To evaluate the end-to-end efficacy of the equalizers, Fig. 4 shows the BER through the entire communication system.

From the simulation results, systems using proposed equalizers benefit from the high reconstruction ability and achieve better BER than conventional LMS. Besides, the ideal case outperforms our proposed approach by taking advantage of the full coding gain, which is reasonable.

Comparing Fig. 3 and Fig. 4, one can notice that the improvement contributed by decoder is smaller than other approaches. However, the proposed equalizer can increase the overall utilization of coding gain, which means the coding gain has been exploited by the NN-based equalizer and the residual coding gain for decoder is limited. The performance gap compared to conventional LMS algorithm is about 0.6dB and 1.5dB for the CNN-based and RNN-based equalizers, respectively.

V. CONCLUSIONS

In this paper, we present a brand new concept to break down the design rule for communication systems. Compared to the conventional block-based design, the proposed CNN-based and RNN-based equalizers not only can mitigate the effects of channel fading, but also can exploit code structure in advance. They therefore increase the overall utilization of coding gain with 0.6 dB and 1.5 dB gain against conventional equalizers.

Our experiment results strongly confirm the fact that convolutional and recurrent neural networks can exploit error correction code structure to become enhanced equalizers, having lower BER compared to their conventional counterparts. However, the early utilization of coding gain may lead to loss in performance of the Viterbi decoder. Therefore, modifications of the decoder are suggested as extension to fully capitalize the gains we have achieved with the NN-based equalizers. Furthermore, the design of encoder may have a significant effect on the performance of the NN-based equalizer, which means that channel-aware encoder design will also be an interesting direction. We believe that our works have opened a brand-new direction of research for channel equalization using NNs.